\documentclass[12pt]{article}%
\usepackage{amsmath}
\usepackage{graphicx}
\usepackage{amsfonts}
\usepackage{amssymb}%
\providecommand{\U}[1]{\protect\rule{.1in}{.1in}}
\begin{document}

\title{Friedel Oscillation about a Friedel-Anderson Impurity}
\author{Yaqi Tao and Gerd Bergmann\\Department of Physics \& Astronomy\\University of Southern California\\Los Angeles, California 90089-0484\\e-mail: bergmann@usc.edu}
\date{\today}
\maketitle

\begin{abstract}
The Friedel oscillations in the vicinity of a Friedel-Anderson (FA) impurity
are investigated numerically. For an FA impurity in the local moment limit the
normalized amplitude $A\left(  \xi\right)  $ is S-shaped, approximately zero
at short distances, approaching two at large distances and crossing the value
one at the characteristic length $\xi_{1/2}$. Surprisingly, the Friedel
oscillations of a simple non-interacting Friedel impurity with a narrow
resonance at the Fermi level show a very similar behavior of their amplitude
$A\left(  \xi\right)  $. A comparison correlates the resonance width and the
Kondo energy of the FA impurity with the characteristic length $\xi_{1/2}$ of
the Friedel oscillations.


\end{abstract}

\section{Introduction}

The properties of magnetic impurities in a metallic host were first studied
theoretically by Friedel \cite{F28} and Anderson \cite{A31}. Kondo \cite{K8}
showed that a magnetic impurity with spin-flip scattering develops a singular
behavior at low temperatures. As a consequence the ground state is a singlet
state with zero effective magnetic moment. Schrieffer and Wolff \cite{S31}
showed that the Friedel-Anderson (FA) Hamiltonian can be transformed into a
Kondo Hamiltonian plus a number of additional terms. Its ground state is also
a singlet state. The disappearance of the magnetic moment at \ low
temperatures, the Kondo effect, is one of the most intensively studied
problems in solid state physics , \cite{D44}, \cite{H23}, \cite{M20},
\cite{A36}, \cite{G24}, \cite{C8}, \cite{H20}, \cite{W18}, \cite{N5},
\cite{N7}, \cite{B103}, \cite{W12}, \cite{A50}, \cite{S29}, \cite{N16}. In the
last decade the Kondo effect has experienced a renaissance. There is a growing
interest in this field \cite{G55}, extending from magnetic atoms on the
surface of corrals \cite{E18} to carbon nanotubes \cite{M93}, quantum dots
\cite{G56}, \cite{A81}, \cite{A80}, \cite{A82}, \cite{B176}, \cite{S83},
\cite{G57} and nanostructures \cite{P41}. There are still many open questions,
particularly the real-space form of the wave function and the resulting charge
density and polarization.

Affleck, Borda and Saleur \cite{A83} (ABS) investigated the formation of
Friedel oscillations in the vicinity of a Kondo impurity. The oscillating part
of the electron charge has the following form.%
\begin{equation}
\rho_{Fr}\left(  r\right)  =\frac{C_{D}}{r^{D}}\left[  A\left(  r\right)
\cos\left(  2k_{F}r+D\frac{\pi}{2}\right)  \right]  \label{fr_osc}%
\end{equation}
where $D$ is the dimension of the system, the coefficients $C_{D}$ have the
values $C_{1}=1/\left(  2\pi\right)  $, $C_{2}=1/\left(  2\pi^{2}\right)  $
and $C_{3}=1/\left(  4\pi^{2}\right)  .$ They succeeded in calculating the
asymptotic form of the function $A\left(  r\right)  $ (which is a universal
funtion of $r/r_{K}$, $r_{K}=\hbar v_{F}/\left(  k_{B}T_{K}\right)  $= Kondo
length) from first principles and performed a numerical calculation using NRG
(numerical renormalization group). $A\left(  r\right)  $ approaches the values
two for $r>>r_{K}$ and zero for $r<<r_{K}$. One of the authors \cite{B178}
applied the approximate solution of the FAIR theory (Friedel Artificially
Inserted Resonance) to reproduce the result by ABS. The two calculations
confirm each other.

In the same paper one of the authors showed that a simple non-interacting
Friedel impurity with its resonance at the Fermi energy showed Friedel
oscillations similar to the Kondo impurity. This reveals the interesting fact
that the complex interacting Kondo impurity yields Friedel oscillations very
similar to those of a simple resonance of a non-interacting electron gas. In
both cases the amplitude $A\left(  r\right)  $ approaches zero for $r<<r_{R}$
and for $r>>r_{R}$ each spin contributes the value one to the amplitude. The
transition happens at a characteristic length which is of the order of
$r_{R}=\hslash v_{F}/\Gamma$ with $v_{F}$ being the Fermi velocity and
$\Gamma$ the half-width of the resonance.

It had been observed earlier that the Friedel phase shifts and sum rules that
were derived for non--interacting electron systems also apply to interacting
electron systems (see for example \cite{A96}). This can be a great help in
describing interacting electron systems, which are generally not very
transparent. A very good example is the Kondo impurity with its Kondo
resonance \cite{G65}, \cite{Z35}, \cite{E17}, \cite{E18}. However, the opinion
of experts (in private conversations) are extremely divided, ranging from
"there is really no Kondo resonance" to " yes there is a Kondo resonance but
its form is very complicated and nobody can write it down". Therefore it is a
considerable help when the complex system follows the same simple rules as a
single electron system. In this paper we calculate the Friedel oscillations of
the Friedel-Anderson (FA) impurity and demonstrate their similarity with the
Friedel oscillations of a simple non-interacting Friedel impurity.

\section{Theoretical Background}

The FA Hamiltonian consists of spin-up and down free electron states
$c_{\nu,\sigma}^{\dagger}$, a d-resonance $d_{\sigma}^{\dagger}$ with the
energy $E_{d}$ and s-d-hopping matrix elements $V_{\nu}^{sd}$ between the
conduction electrons and the d-impurity. A simultaneous occupation
$n_{d\uparrow}$ and $n_{d\downarrow}$ of the d-resonance contributes a Coulomb
exchange energy $Un_{d\uparrow}n_{d\downarrow}.$%

\begin{equation}
H_{FA}=%
{\textstyle\sum_{\sigma}}
\left\{  \sum_{\nu=0}^{N-1}\varepsilon_{\nu}c_{\nu\sigma}^{\dag}c_{\nu\sigma
}+\sum_{\nu=0}^{N-1}V_{\nu}^{sd}[d_{\sigma}^{\dag}c_{\nu\sigma}+c_{\nu\sigma
}^{\dag}d_{\sigma}]+E_{d}d_{\sigma}^{\dagger}d_{\sigma}\right\}
+Un_{d\uparrow}n_{d\downarrow} \label{H_FA}%
\end{equation}

The FA impurity shows also the Kondo effect as Schrieffer and Wolff \cite{S31}
have shown. The FA impurity has many degrees of freedom. Its behavior is
determined by the d-resonance energy $E_{d}$, the Coulomb energy $U$ and the
s-d-scattering matrix element $\left\vert V_{sd}\right\vert ^{2}$. For
simplicity we will discuss here the symmetric FA impurity with $E_{d}=-U/2$.

For sufficiently large values of $U/\left\vert V_{sd}\right\vert ^{2}$ the FA
impurity possesses a magnetic moment above the Kondo temperature. One
difficulty in determining the magnetic moment is the fact that the magnetic
state is not the ground state of the system. The ground state is a singlet
state (or Kondo state) with vanishing the total moment, whose energy is
lowered by the Kondo energy. For many realistic systems the Kondo energy is
quite small. If warmed above the Kondo temperature the FA impurity is
magnetic, but it is still at rather low temperature. We denote this as the
magnetic pseudo-ground state. In the FAIR approach the magnetic wave function
has the form
\begin{equation}
\Psi_{MS}=\left[  Aa_{0\uparrow}^{\dagger}b_{0\downarrow}^{\dagger
}+Ba_{0\uparrow}^{\dagger}d_{\downarrow}^{\dagger}+Cd_{\uparrow}^{\dagger
}b_{0\downarrow}^{\dagger}+Dd_{\uparrow}^{\dagger}d_{\downarrow}^{\dagger
}\right]  \prod_{i=1}^{n-1}a_{i\uparrow}^{\dagger}\prod_{i=1}^{n-1}%
b_{i\downarrow}^{\dagger}\Phi_{0} \label{Psi_MS}%
\end{equation}

Here $a_{0\uparrow}^{\dagger}$ and $b_{0\downarrow}^{\dagger}$ are two
artificially inserted Friedel resonances with a composition
\[
a_{0\uparrow}^{\dagger}=%
{\displaystyle\sum_{\nu}}
\alpha_{0}^{\nu}c_{\nu\uparrow}^{\dagger}%
\]
(As discussed in previous papers \cite{B93} the FAIR states have a relatively
simple interpretation in Hilbert space.) The remaining states in the spin up
basis are $\left\{  a_{i\uparrow}^{\dagger}\right\}  $ which are orthogonal to
each other and to $a_{0\uparrow}^{\dagger}.$In addition the free electron
Hamiltonian $H_{0\uparrow}$ for spin-up electrons is made sub-diagonal in this
basis, i.e. all matrix elements $\left\langle a_{\nu\uparrow}^{\dagger
}\left\vert H_{0\uparrow}\right\vert a_{\mu\uparrow}^{\dagger}\right\rangle
$=$\delta_{\nu,\mu}$ for $\nu,\mu\neq0$. As a consequence the whole basis is
fully determined by $a_{0\uparrow}^{\dagger}$. The FAIR state $b_{0\downarrow
}^{\dagger}$ with its basis $\left\{  b_{i\downarrow}^{\dagger}\right\}  $ has
analogous properties. The details of the FAIR theory are described in previous
papers \cite{B151}, \cite{B152}, \cite{B153}, etc.

The mean field theory produces two d-resonances in the magnetic state, one for
the spin-up band at $E_{d}+Un_{d\downarrow}$ and another one for the spin-down
band at $E_{d}+Un_{d\uparrow}.$ Here $n_{d\uparrow},n_{d\downarrow}$ are the
partial occupations of the spin-up and -down d-resonances. In mean field
theory these resonances have the half width $\Gamma=\pi\left\vert
V_{sd}\right\vert ^{2}\rho$ ($\rho$=density of states). Mean field theory
completely decouples the s-d-scattering between different spin bands. In
reality a doubly occupied d-state decays in both spin bands, and the decay
rate is twice as large \cite{L57}. The FAIR theory of the magnetic state
yields d-resonances with the correct resonance width \cite{B181}.

The ground state of the FA impurity is a singlet state. The latter is
obtained\ in FAIR by reversing all spins in equ. (\ref{Psi_MS}). After
ordering the spin sequences the two states are added and normalized. For the
singlet state the composition of the FAIR states $a_{0}^{\dagger}$ and
$b_{0}^{\dagger}$ has to be optimized again (and is very different from the
composition in the magnetic state). The two (opposite) magnetic states in
(\ref{Psi_SS}) are not orthogonal to each other and yield a finite
interference between the two magnetic components. The strength of this
interference determines the Kondo energy.%

\begin{align}
\Psi_{SS}  &  =\label{Psi_SS}\\
&  \left\{
\begin{array}
[c]{c}%
\left[  Aa_{0\uparrow}^{\dagger}b_{0\downarrow}^{\dagger}+Ba_{0\uparrow
}^{\dagger}d_{\downarrow}^{\dagger}+Cd_{\uparrow}^{\dagger}b_{0\downarrow
}^{\dagger}+Dd_{\uparrow}^{\dagger}d_{\downarrow}^{\dagger}\right]
\prod_{i=1}^{n-1}a_{i\uparrow}^{\dagger}\prod_{i=1}^{n-1}b_{i\downarrow
}^{\dagger}\Phi_{0}\\
+\left[  A^{\prime}b_{0\uparrow}^{\dagger}a_{0\downarrow}^{\dagger}+B^{\prime
}d_{\uparrow}^{\dagger}a_{0\downarrow}^{\dagger}+C^{\prime}b_{0\uparrow
}^{\dagger}d_{\downarrow}^{\dagger}+D^{\prime}d_{\uparrow}^{\dagger
}d_{\downarrow}^{\dagger}\right]  \prod_{i=1}^{n-1}b_{i\uparrow}^{\dagger
}\prod_{i=1}^{n-1}a_{i\downarrow}^{\dagger}\Phi_{0}%
\end{array}
\right\} \nonumber
\end{align}%
\[
\]

In the numerical calculation we use the Wilson band \cite{W18} which is an
electron-hole symmetric band, normalized by the Fermi energy and extending
from $\left(  -1<\varepsilon<+1\right)  $ with a constant density of states
$\rho=1/2$. The electron states in the Wilson band are represented by a finite
number of Wilson states. We generally start our numerical calculation with
$N=60$ Wilson states with energies of $\pm\frac{3}{4},\pm\frac{3}{8},\pm
\frac{3}{16},$.. $\pm\frac{3}{2^{30}},\pm\frac{1}{2^{30}}$. The ratio between
neighboring energies is $\Lambda=2$ (except for the two states next to the
Fermi level). The wave function of the Wilson states in real space is
described in previous papers \cite{B177}, \cite{B178}. From the composition of
the two bases $\left\{  a_{i}^{\dagger}\right\}  $ and $\left\{
b_{i}^{\dagger}\right\}  $ in terms of the Wilson states $c_{\nu}^{\dagger}$
one obtains their wave functions in real space.

The calculation in real space can be done in one, two or three dimensions. The
results differ by the dimensional factor $C_{D}/r^{D}$and the phase shift
$-D\frac{\pi}{2}$ (see equ.(\ref{fr_osc})). If one splits off the dimensional
factor then we expect that the amplitude $A\left(  r\right)  $ is the same for
all dimensions. Therefore we choose the simplest case, the one-dimensional FA impurity.

The calculation of the real space density at the position $r$ is in principle
straight forward but rather tedious due to the interference terms between the
different bases. Here one has to calculate a large number of multi-electron
scalar products in form of determinants. Fortunately this has to be done only
once and can be used for all distances.

\section{Numerical Results and Discussion}

We performed the calculation of the Friedel oscillation for different
combinations of parameters. In a first series\ we used for the s-d-hopping
matrix element the value $\left\vert V_{sd}\right\vert ^{2}$ $=0.03$ and for
the Coulomb energy the values $U=0.1,$ $0.2$, $0.4$, $0.6$, $0.8$ and $1.0$
with $E_{d}=-U/2$. In the first round $60$ Wilson states with energies of
$\pm\frac{3}{4},\pm\frac{3}{8},\pm\frac{3}{16},$.. $\pm\frac{3}{2^{30}}%
,\frac{1}{2^{30}}$ represented the free electron basis. As ABS already noticed
the Wilson states with an energy ratio of $\Lambda=2$ are so far apart in
their energy that the results are not yet very accurate. (For example the
Wilson state with the energy $\frac{3}{4}$ is composed of all states in the
energy range of $\frac{1}{2}<\varepsilon<1.$This means that it has roughly an
energy uncertainty of $\frac{1}{4}$. In simple words, the energy uncertainty
of the Wilson states divided by the energy $\Delta E_{W}/E_{W}$ is about $1/3$
for $\Lambda=2$). In the FAIR approach it is relatively easy to sub-divide the
Wilson states geometrically \cite{B178}. The FAIR states $a_{0}^{\dagger}$ and
$b_{0}^{\dagger}$ can be well interpolated during this procedure. From the
FAIR states one obtains the whole bases $\left\{  a_{i}^{\dagger}\right\}  $
and $\left\{  b_{i}^{\dagger}\right\}  $.

After the first sub-division we have a new Wilson basis of 120 states with an
energy ratio of $\Lambda=2^{1/2}$. The second sub-division yields 240 states
with $\Lambda=2^{1/4}\approx\allowbreak1.\,\allowbreak19$. Now even the large
energy levels lie close together, and the energy uncertainty $\Delta
E_{W}/E_{W}$ takes the value of about $\allowbreak0.09$.

In the calculation we measure the distance from the impurity in units of half
the Fermi wave length $\lambda_{F}$, i.e. $\xi=2r/\lambda_{F}$. In these units
the Friedel oscillations have the period "one". We multiply the actually
calculated amplitude by the factor $2\pi\xi$ to cancel the one-dimensional
prefactor in equ. (\ref{fr_osc}). The resulting normalized amplitude $A\left(
\xi\right)  $ is then valid for all dimensions of the sample.

In Fig.1 the Friedel oscillation amplitude $A\left(  \xi\right)  $ is shown as
the function of the logarithm of the distance $\ell=\log_{2}\left(
\xi\right)  $ for the different parameters.%

\begin{align*}
&
{\includegraphics[
height=3.4271in,
width=4.0598in
]%
{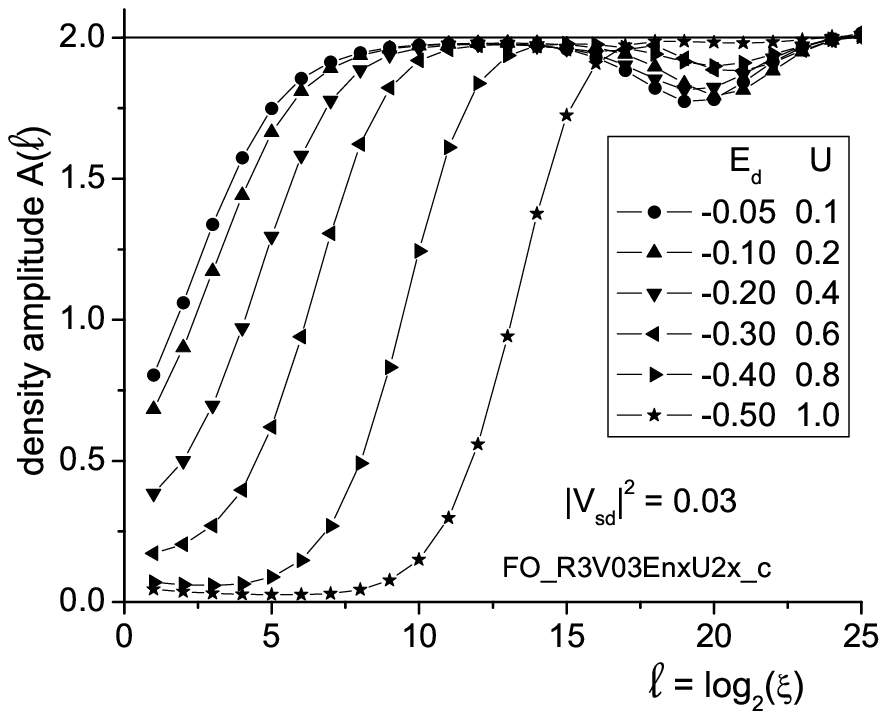}%
}%
\\
&
\begin{tabular}
[c]{l}%
Fig.1: The amplitude $A\left(  \ell\right)  $ of the Friedel oscillations for
a symmetric\\
FA impurity with $\left\vert V_{sd}\right\vert ^{2}=0.03$ and different values
of $U$ with $E_{d}=-U/2$.\\
The abscissa is the distance $\ell=\log_{2}\left(  \xi\right)  $ from the
impurity where $\xi=2r/\lambda_{F}.$%
\end{tabular}
\end{align*}

The right curve (stars) in Fig.1 has the largest $U=1$ and behaves very
similar to a Kondo impurity that is discussed in ref. \cite{A83}, \cite{B178}.
For large distances the amplitude $A\left(  \xi\right)  $ reaches the value
$2,$ and for short distances the amplitude essentially vanishes. The
transition between these two regions occurs at the $\ell_{1/2}$-point where
$A\left(  \ell\right)  =1$. For $U=0.8$ one obtains a similar curve as for
$U=1.0$, only shifted to the left. With decreasing $U$ the $\ell_{1/2}$-point
for the amplitude moves to shorter distances

For $U=1$ and $U=0.8$ the positions of the $\ell_{1/2}$-points are at $9.4$
and $13.1$. If one shifts one of the curves by $3.\,\allowbreak7$ then one
obtains a single curve. It represents the universal curve for the FA impurity
with a full magnetic moment. If $U$ is further decreased then the curves for
$A\left(  \xi\right)  $ are further shifted to the left and for $U\leq0.4$
they don't reach the horizontal axis anymore. Essentially the whole
environment of the impurity is filled with Friedel oscillations. Surprisingly
the curves develop a small minimum at about $\ell=20$ with decreasing value of
$U$ .

For the smallest value of $U=0.1$ one is in the perturbative region while for
$U=1.0$ one is in the local moment region (see for example Krishna-murtha et
al. \cite{K58}). Generally the properties of the symmetric FA impurity are
characterized by the parameter $U/\left(  \pi\Gamma\right)  =U/\left(  \pi
^{2}\left\vert V_{sd}\right\vert ^{2}\rho\right)  $. In table I the values of
$U/\left(  \pi\Gamma\right)  $ are collected for the different examples. Our
curves for the Friedel oscillations approach the local moment behavior
somewhere above $U/\left(  \pi\Gamma\right)  \approx4$.
\begin{align*}
&
\begin{tabular}
[c]{|l|l|l|l|l|}\hline
\textbf{U} & $\left\vert V_{sd}\right\vert ^{2}$ & $\mathbf{U/\pi\Gamma}$ &
$\mathbf{r}$ & \textbf{S}\\\hline
0.1 & 0.03 & $0.66$ & 0.51 & 0.303\\\hline
0.2 & 0.03 & $1.\,\allowbreak33$ & 0.67 & 0.354\\\hline
0.4 & 0.03 & $2.\,\allowbreak67$ & 0.83 & 0.434\\\hline
0.6 & 0.03 & $4.05$ & 0.90 & 0.479\\\hline
0.8 & 0.03 & $5.\,\allowbreak33$ & 0.93 & 0.497\\\hline
1.0 & 0.03 & $6.\,\allowbreak67$ & 0.95 & 0.500\\\hline
\end{tabular}
\\
&
\begin{tabular}
[c]{l}%
Table I: The effective coupling strength $U/\left(  \pi\Gamma\right)  $ for
the different\\
parameter of Fig.1. $r$ and $S$ are discussed in the text.
\end{tabular}
\end{align*}

\subsection{The magnetic half of the singlet state}

The FAIR solution for the FA-impurity consists of two entangled magnetic
states as shown in equ. (\ref{Psi_SS}). Each half in equ. (\ref{Psi_SS})
possesses a magnetic moment. For large $U/\left(  \pi\Gamma\right)  $ the
magnetic moment is relatively well defined because the sum $S=\left(
A^{2}+B^{2}+C^{2}+D^{2}\right)  $ is close to 0.5 and almost normalized to
$1/2$. If we define $r=\left(  B^{2}-C^{2}\right)  /\left(  A^{2}+B^{2}%
+C^{2}+D^{2}\right)  $ then the moment has the value $\mu/\mu_{B}\approx r$ as
long as $S\approx0.5.$ However, with decreasing $U/\left(  \pi\Gamma\right)  $
the sum $S$ becomes considerably smaller than $1/2$ because it is the singlet
state that is normalized, and the interference between the two magnetic parts
becomes more and more important. The values of $r$ and $S$ are also collected
in table I. It should, however, be emphasized that the magnetic moment of the
magnetic half in the singlet state is not the same as the magnetic moment of
the corresponding magnetic state as presented by equ. (\ref{Psi_MS}). Although
the structure of the states is the same, the FAIR states $a_{0}^{\dagger}$ and
$b_{0}^{\dagger}$ assume rather different composition.

For $U=1,E_{d}=-0.5$ and $\left\vert V_{sd}\right\vert ^{2}=0.03$ the magnetic
halves of the singlet state are very well normalized to the value of
$S=0.500$. Therefore it is interesting to study their properties. In Fig.2 the
half of the ground state with the net d-spin down is investigated. Both the
spin-up and the spin-down bands of the conduction electron gas show charge
oscillations with the period one. Their magnitudes are identical within the
accuracy of the calculation. At large distances they reach the value $A=0.5$.
This means that the total charge oscillation of both spins for both magnetic
components yields the value two which agrees with the results of Fig.1.
However, for short distances one observes for each spin band a value of 0.05.
This value appears to contradict the vanishing Friedel oscillations in Fig.1.%

\begin{align*}
&
{\includegraphics[
height=3.2893in,
width=4.0473in
]%
{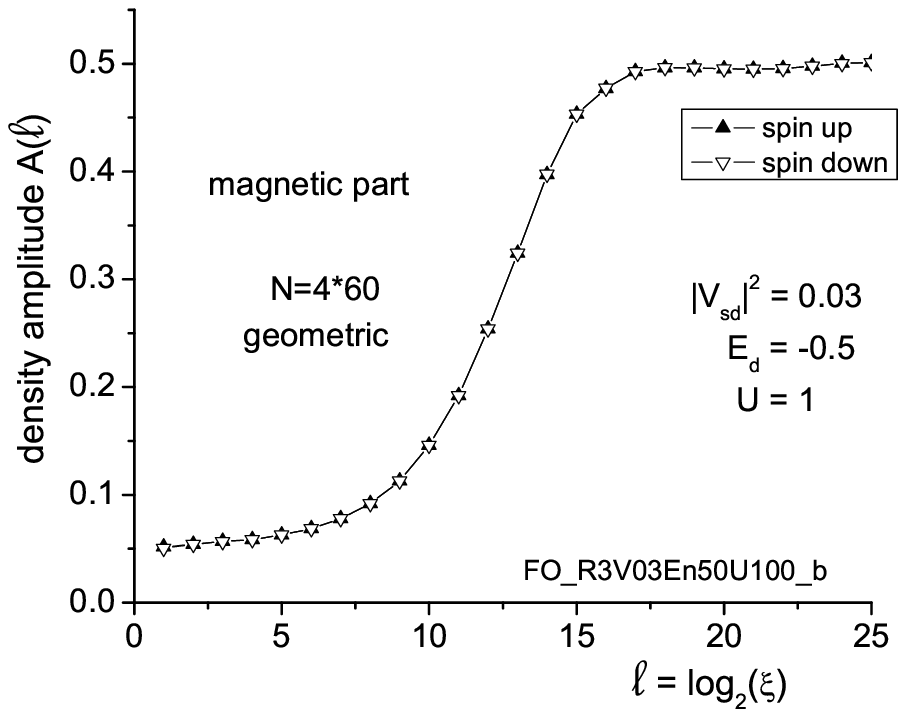}%
}%
\\
&
\begin{tabular}
[c]{l}%
Fig.2: The amplitude $A\left(  \xi\right)  $ of the charge oscillations of the
spin-up (full\\
triangle up) and the spin-down (open triangle down) electrons in the\\
magnetic component of the singlet state.
\end{tabular}
\end{align*}

The reason for this apparent contradiction lies in the phase of the different
oscillations. The position of the maxima of the charge oscillations may be at
$\xi_{ma}$. In the absence of a phase shift the maxima of the Friedel
oscillations would be at integer values of $\xi$. Therefore the deviation of
the $\xi_{ma}$ from an integer, $\delta\xi,$ yields the phase shift in units
of $2\pi$. The positions of $\delta\xi$of the maxima and minima of the charge
oscillations of the spin-up part and the spin-down part of the conduction
electrons are plotted in Fig.3. The full triangles show $\delta\xi_{mi}$ of
the minima of the charge oscillations and the open triangles represent the
maxima. In addition the up-triangles represent the spin-up electrons and the
down-triangles the spin-down electrons.

For large distances the maxima positions for both spin orientations lie at
0.75 and the minima at 0.25. Therefore the charge oscillations add
constructively in this region. However for distances which are smaller than
$\xi_{1/2}$ the maxima and minima of spin-up and down electrons are shifted by
$\pm0.25$ yielding a relative difference of $\delta\xi=0.5$, corresponding to
a phase difference of $\pi$. Therefore the charge oscillations of spin-up and
down electrons cancel in this region.
\begin{align*}
&
{\includegraphics[
height=4.2266in,
width=4.9904in
]%
{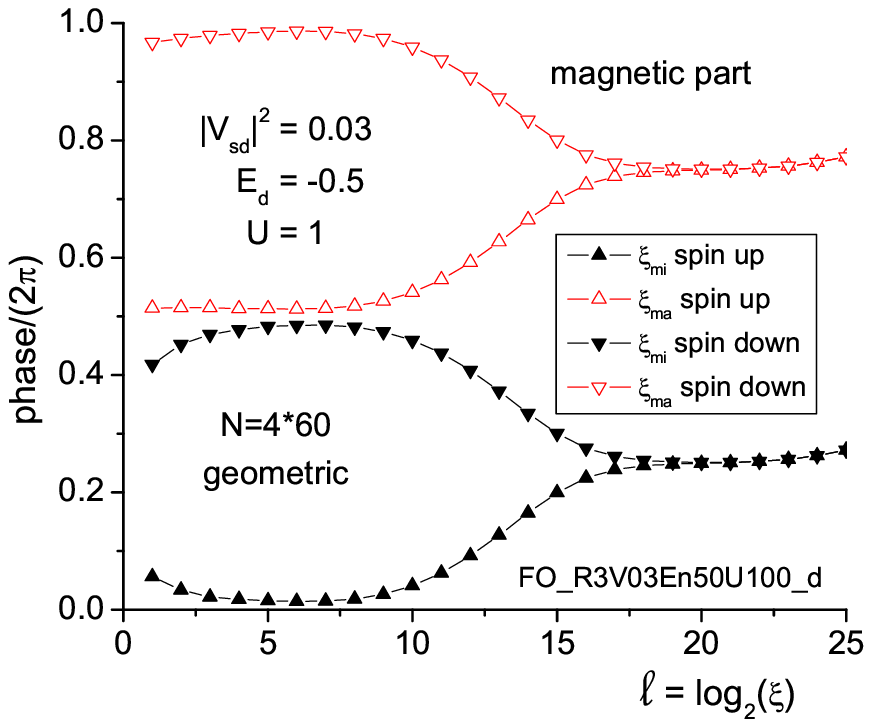}%
}%
\\
&
\begin{tabular}
[c]{l}%
Fig.3: The phase relations of the charge oscillations of the magnetic half
of\\
the FA impurity with $U=1,E_{d}=-0.5$ and $\left\vert V_{sd}\right\vert
^{2}=0.03$ as a function of the\\
distance $\ell=\log_{2}\left(  \xi\right)  $. The full and open triangles show
the positions (modulo one) of the\\
minima and maxima of the Friedel oscillations for the spin-up and spin-down\\
electron (up and down triangles).
\end{tabular}
\end{align*}

While the total charge oscillation in the magnetic half of the singlet
solution cancels essentially to zero in the region of $\ell<<\ell_{1/2}$ one
obtains a polarization oscillation in this region. In Fig.4 the amplitude of
this polarization in plotted as a function of $\ell=\log_{2}\left(
\xi\right)  $.
\begin{align*}
&
{\includegraphics[
height=3.3566in,
width=4.0747in
]%
{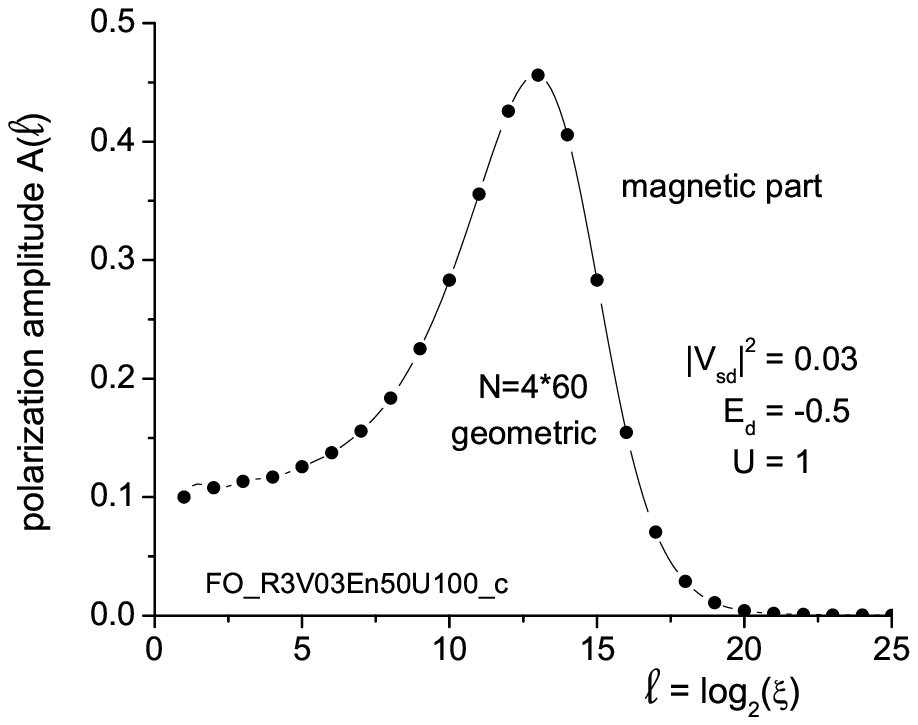}%
}%
\\
&
\begin{tabular}
[c]{l}%
Fig.4: The amplitude of the polarization oscillation of the\\
magnetic magnetic component for the $U=1$ FA impurity\\
as a function of the distance $\ell=\log_{2}\xi$.
\end{tabular}
\end{align*}

If one performs the same analysis for the FA impurity with $U=0.1,E_{d}=-0.05$
and $\left\vert V_{sd}\right\vert ^{2}=0.03$ one observes essentially two
important differences. Fig.5 shows that (i) the phase difference between
spin-up and down electrons at short distances is much less developed than in
the local moment example in Fig.3, and (ii) the transitional region where the
phases split is shifted to much shorter distances. The fact that the phase
differences at short distances is much less than $\pi$ is the reason that the
two charge oscillations of spin-up and down electrons do not cancel. The
Friedel oscillations maintain a relatively large amplitude even at short distances.%

\begin{align*}
&
{\includegraphics[
height=3.3441in,
width=3.9493in
]%
{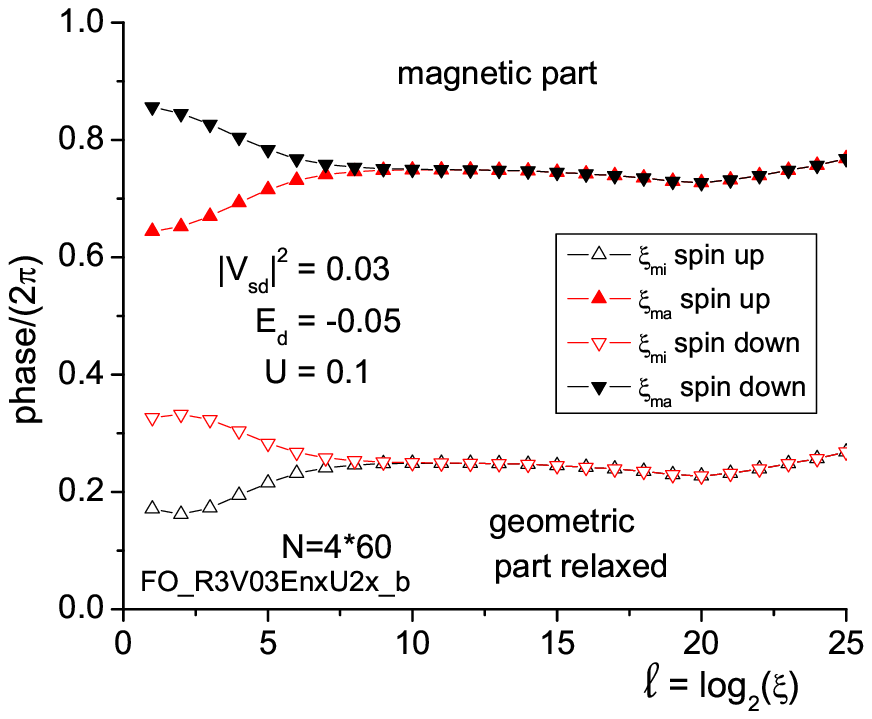}%
}%
\\
&
\begin{tabular}
[c]{l}%
Fig.5: The phase relations of the charge oscillations of the magnetic half
of\\
the FA impurity with $U=0.1,E_{d}=-0.05$ and $\left\vert V_{sd}\right\vert
^{2}=0.03$ as a function of the\\
distance $\xi$. The full and open triangles show the positions (modulo one) of
the\\
minima and maxima of the Friedel oscillations for the spin-up and spin-down\\
electron (up and down triangles).
\end{tabular}
\end{align*}

\subsection{Fermi liquid picture}

Nozieres \cite{N14} derived from Wilson's renormalization \cite{W18} and
Anderson's scaling theory \cite{A95} a Fermi liquid description of the Kondo
effect at low temperatures. In this description, which should also apply to
the FA impurity in the local moment regime, the magnetic moment forms a
strongly bound singlet state with a conduction electron. This removes the
magnetic moment as well as one conduction electron from the system. Virtual
excitations cause a weak interacton between the quasi-particles. In ref
\cite{B178} one of the authors pointed out the similarity between the Friedel
oscillations of a Kondo impurity and a Friedel impurity with a very narrow
resonance at the Fermi level. This is another confirmation of the Fermi liquid
description of a Kondo impurity. In the present paper we want to show that
this simililarity also extends to the FA impurity.

Fig.6 shows the Friedel oscillations of a simple Friedel impurity with spin-up
and down sub-bands. The resonance d-state lies at the Fermi level, i.e.
$E_{d}=0$. The s-d-hopping matrix element $\left\vert V_{sd}\right\vert ^{2}$
is varied by four orders of magnitude between $10^{-1}$ and $10^{-5}$.

There is a remarkable similarity between the Friedel oscillations of the FA
impurity in Fig.1 and of the Friedel impurity in Fig.6. If we consider first
the Friedel impurity with $\left\vert V_{sd}\right\vert ^{2}=10^{-5}$ then the
overall shape of the $A\left(  \ell\right)  $ curve is very similar to the
$A\left(  \ell\right)  $-curve for the FA impurity with $U=1$. Both curves
approach the value two at large distances and zero at short distances. For the
Friedel impurity an increase in $\left\vert V_{sd}\right\vert ^{2}$ by a
factor of 10 shifts the curve by $\log_{2}\left(  10\right)  $ to the left.
The two left curves in Fig.6 no longer reach the value zero. This must be due
to the fact that $\Gamma$ is no longer sufficiently small compared to the band
width of one.

We observed a similar behavior for the FA impurity when we reduced the value
of $U$. The similarity with the Friedel impurity suggests that the reason here
is also due to the finite band width and not due to the fact that the FA
impurity with $U/\left(  \pi\Gamma\right)  <3$ is no longer in the local
moment limit.

For the Friedel impurity with $\left\vert V_{sd}\right\vert ^{2}=10^{-5}$ the
position of the $\ell_{1/2}$ point is at $\ell_{1/2}\approx12.6$,
corresponding to a length of $\xi_{1/2}=2^{12.6}\approx$ $6208.$ The product
between $\xi_{1/2}$ and the resonance half-width $\Gamma=\pi\left\vert
V_{sd}\right\vert ^{2}\frac{1}{2}=1.\,\allowbreak57\times10^{-5}$ is
$\xi_{1/2}\Gamma\approx1.\,\allowbreak57\times10^{-5}\ast2^{12.6}\approx0.10$
($\Gamma$ is measured in units of the Fermi energy $E_{F}$ and $\xi$ in units
of half the Fermi wave length $\lambda_{F}/2$). This product is universal (as
long as $\Gamma$ is sufficiently small compared with band width of the system).%

\begin{align*}
&
{\includegraphics[
height=3.2744in,
width=3.9767in
]%
{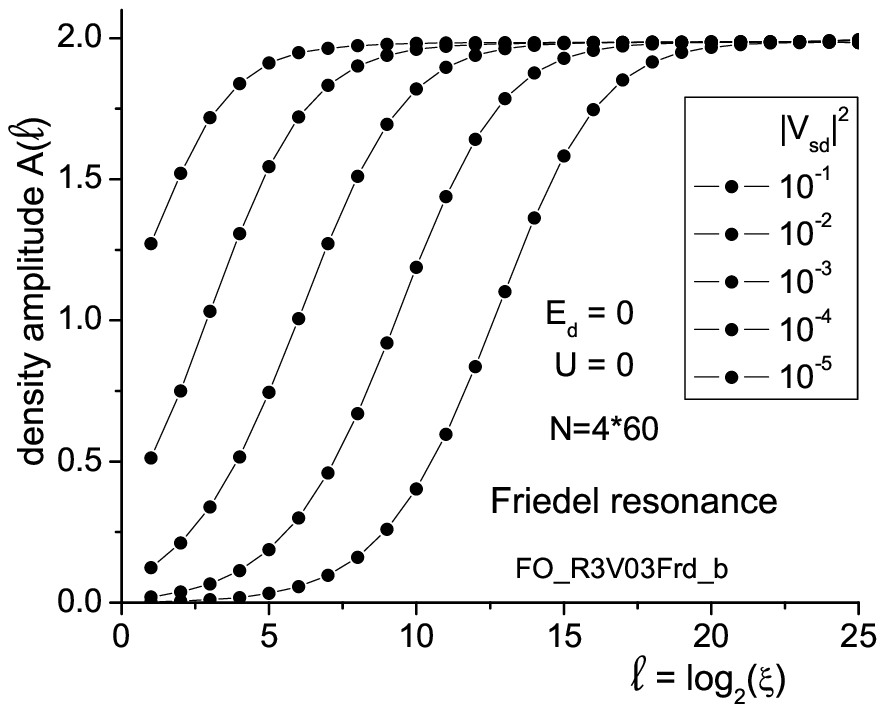}%
}%
\\
&
\begin{tabular}
[c]{l}%
Fig.6: The amplitudes of the Friedel oscillations for different\\
Friedel resonances with $E_{d}=0$ and $\left\vert V_{sd}\right\vert
^{2}=10^{-j}$ with $j=1,2,3,4,5$.
\end{tabular}
\end{align*}

We want to quantify this similarity. For this purpose we determine for each
curve in Fig.1 and Fig.6 the distance $\ell_{1/2}=\log_{2}\left(  \xi
_{1/2}\right)  $ where the amplitude $A\left(  \ell\right)  $ has half the
saturation value $A\left(  \ell_{1/2}\right)  =1$. For the Friedel impurity we
plot $\log_{2}\Gamma$ of the resonance half-width $\Gamma$ as a function of
$\ell_{1/2}=\log_{2}\xi_{1/2}$. That yields the full points in Fig.7.

The similarity between the Friedel curves in Fig.6 and the FA curves in Fig.1
supports the interpretation that the FA impurity has a resonance at the Fermi
energy. However, its half-width is not well known. We assume that it is
proportional to the Kondo energy (which is differently defined in different
theoretical approaches). We choose for the FA impurity in Fig.1 the definition
of the Kondo energy by means of the susceptibility. The latter we obtain
through a linear response calculation \cite{B182} yielding $E_{\chi}$. In
Fig.7 we plot the logarithm of this Kondo energy $\log_{2}\left(  E_{\chi
}\right)  $ versus $\ell_{1/2}=\log_{2}\left(  \xi_{1/2}\right)  $ for each
curve in Fig.1 (stars). Both sets of points for the FA and the Friedel
impurity (circles and stars) lie on two straight lines with the slope of $-1$
and are separated by a vertical distance of 1.25 which corresponds to a factor
of 2.4. If we assign to the FA impurity a resonance half-width of $\Gamma
_{FA}\approx E_{\chi}/2.4$ then the values for $\Gamma_{F}$ and $\Gamma_{FA}$
fall on the same straight line with the slope of $-1$.

Our calculations of the Friedel oscillations not only show that there is a
Kondo resonance at the Fermi level but also suggest that even for relatively
small $U/\left(  \pi\Gamma\right)  $ the quasi-particle behavior is similar to
that of a Friedel impurity.%

\begin{align*}
&
{\includegraphics[
height=3.1905in,
width=3.9344in
]%
{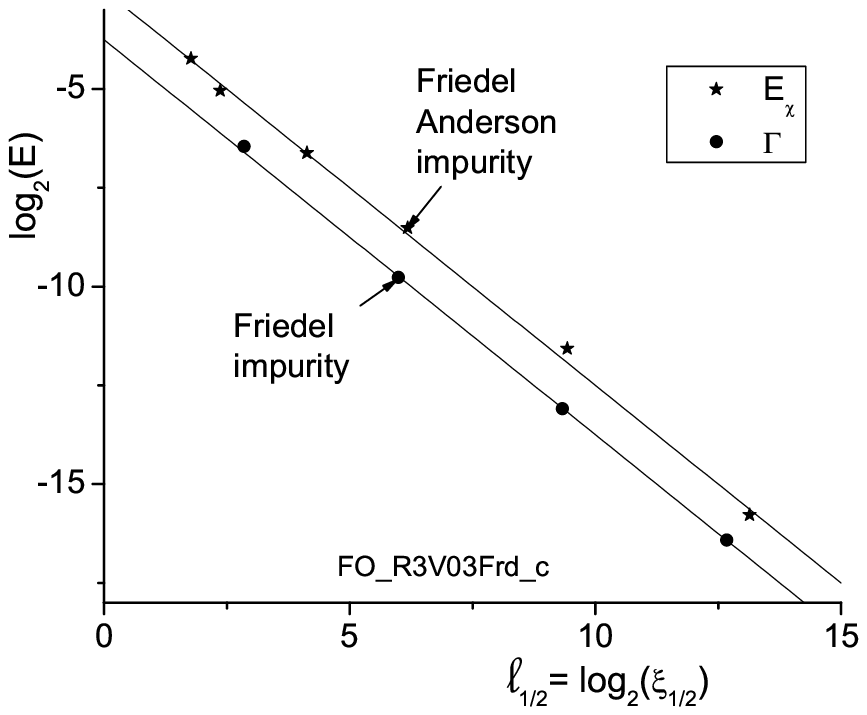}%
}%
\\
&
\begin{tabular}
[c]{l}%
Fig.7: The logarithm of the half-width $\Gamma_{F}$ for the Friedel impurity
(full circles)\\
and the logarithm of the Kondo energy for the FA impurity are plotted\\
versus the corresponding values $\ell_{1/2}=\log_{2}\left(  \xi_{1/2}\right)
$-values. The latter are obtained\\
from Fig.6 and Fig.1 as the position where $A\left(  \ell\right)  $ takes the
value $A\left(  \ell_{1/2}\right)  =1$. The\\
two full lines are straight lines with the slope minus one. Their vertical\\
separation is $1.25$, corresponding to a ratio of $2^{1.25}\approx$ $2.\,4$
between Kondo\\
energy and Kondo resonance half width.
\end{tabular}
\end{align*}

\section{Conclusion}

In this paper we investigate the Friedel oscillations of a set of symmetric
Friedel-Anderson impurities. In the local moment limit (large $U/\pi\Gamma$)
the normalized amplitude $A\left(  \xi\right)  $ essentially vanishes for
short distances and assumes the value 2 for large distances. In this range of
$U/\left(  \pi\Gamma\right)  $ the $A\left(  \xi\right)  $-curves show
universal behavior because curves for different $U/\left(  \pi\Gamma\right)  $
can be shifted into a perfect overlap. The physics behind this behavior of the
amplitude of the Friedel oscillation is illuminated by the study of the simple
Friedel impurity with a very narrow resonance at the Fermi level. For these
non-interacting Friedel impurities one obtains real space Friedel oscillations
which are very similar to those of the interacting FA impurities.

In the case of the Friedel impurity one can derive two conclusions from the
Friedel oscillations:

\begin{itemize}
\item The amplitude $A\left(  \xi\right)  $ reaches the saturation value of
two only when the resonance lies exactly at the Fermi level.

\item The $\xi_{1/2}$-point where the normalized amplitude $A\left(  \xi
_{1/2}\right)  $ is equal to one yields the resonance half-width $\Gamma_{F}$
through the universal relation $\Gamma_{F}\xi_{1/2}\approx0.10$.
\end{itemize}

The similarity between the Friedel oscillations of the FA and the Friedel
impurity supports the concept of a "Kondo" resonance at the Fermi level. The
position of the $\xi_{1/2}$-point suggests that the half-width of the FA
resonance is of the order of the Kondo energy $\Gamma_{FA}\approx E_{\chi
}/2.4$ .

Our calculations of the Friedel oscillations support the concept that there is
a Kondo resonance at the Fermi level with the phase shift $\pi/2$ and suggests
that even for relatively small values of $U/\left(  \pi\Gamma\right)  $ the
quasi-particle behavior is similar to that of a Friedel impurity.

It should be emphasized that the spatial behavior of the FAIR wave function
and its charge and spin oscillations provides quantitative information about
Kondo energy, Kondo length and resonance width of a FA impurity without the
need of a magnetic field or excited states.


\newpage

\begin{thebibliography}{99}                                                                                               %


\bibitem {F28}J. Friedel, Adv. Phys. 3, 446 (1954);Can. J. Phys. 34, 1190
(1956); Nuovo Cimento Suppl. 7, 287 (1958); J. Phys. Radium 19, 573 (1958)

\bibitem {A31}P. W. Anderson, Phys. Rev. 124, 41 (1961)

\bibitem {K8}J. Kondo, Prog. Theor. Phys. 32, 37 (1964)

\bibitem {S31}J. R. Schrieffer and P. A. Wolff, Phys. Rev. 149, 491 (1967)

\bibitem {D44}M. D. Daybell, and W. A. Steyert, Rev. Mod. Phys. 40, 380 (1968)

\bibitem {H23}A. J. Heeger, in Solid State Physics, ed. by F. Seitz, D.
Turnbull, and H. Ehrenreich (Academic, New York, 1969), Vol 23, p284

\bibitem {M20}M. B. Maple, in "Magnetism", edited by G. T. Rado and H. Suhl
(Academic, New York, 1973), Vol. V, p. 289

\bibitem {A36}P. W. Anderson, Rev. Mod. Phys. 50, 191 (1978)

\bibitem {G24}G. Gruener and A. Zavadowski, Prog. Low Temp. Phys. 7B, 591 (1978)

\bibitem {C8}P. Coleman, J. Magn. Magn. Mat. 47, 323 (1985)

\bibitem {H20}A. C. Hewson, The Kondo problem to heavy Fermions, Cambridge
University Press, 1993

\bibitem {W18}K. G. Wilson, Rev. Mod. Phys. 47, 773 (1975)

\bibitem {N5}P. Nozieres, Ann. Phys. (Paris) 10, 19 (1985)

\bibitem {N7}D. M. Newns and N. Read, Adv. in Phys. 36, 799 (1987)

\bibitem {B103}N. E. Bickers, Rev. Mod. Phys. 59, 845 (1987)

\bibitem {W12}P. B. Wiegmann, in Quantum Theory of Solids, edited by I. M.
Lifshitz (MIR Publishers, Moscow, 1982), p. 238

\bibitem {A50}N. Andrei, K. Furuya, and J. H. Lowenstein, Rev. Mod. Phys. 55,
331 (1983)

\bibitem {S29}P. Schlottmann, Phys. Reports 181, 1 (1989)

\bibitem {N16}P. Nozieres, and A. Blandin, J. Physique 41, 193 (1980)

\bibitem {G55}L. Kouwenhoven, and L. Glazman, Phys. World 14, 33 (2001)

\bibitem {E18}H. C. Manoharan, C. P. Lutz and D. M. Eigler, Nature (London)
403, 512 (2000)

\bibitem {M93}J. Paaske, A. Rosch, P. Woelfle, N. Mason, C. M. Marcus, J.
Nygard, Nature Physics, 2, 460 (2006)

\bibitem {G56}M. G. Vavilov and L. I. Glazman, Phys. Rev. Lett. 94, 086805 (2005)

\bibitem {A81}I. Affleck, and P. Simon, Phys. Rev. Lett. 86, 2854 (2001)

\bibitem {A80}I. P. Simon and I. Affleck, Phys. Rev. Lett. 89, 206602 (2002)

\bibitem {A82}R. G. Pereira, N. Laflorencie, I. Affleck, and B. I. Halperin,
arXiv:cond-mat/0612635 (2007)

\bibitem {B176}L. Borda, Phys. Rev. B 75, 041307(R) (2007)

\bibitem {S83}J. Simonin, arXiv:0708. 3604 (2007)

\bibitem {G57}I. L. Aleiner, P. W. Brouwer and L. I. Glazman, Phys. Rep. 358,
309 (2002)

\bibitem {P41}M. Pustilnik, physica stat. solidi (a) 203 1137 (2006)

\bibitem {A83}I. Affleck, L. Borda, H. Saleur, Phys. Rev. B 77, 180404(R) (2008)

\bibitem {B178}G. Bergmann, Phys. Rev. B 78, 195124 (2008)

\bibitem {B93}G. Bergmann, physics today, 32, 25 (August 1979)

\bibitem {B151}G. Bergmann, Phys. Rev. B 74, 144420 (2006)

\bibitem {B152}G. Bergmann, Phys. Rev. B 73, 092418 (2006)

\bibitem {B153}G. Bergmann and L. Zhang, Phys. Rev. B 76, 064401 (2007)

\bibitem {L57}D. E. Logan, M. P. Eastwood, and M. A. Tusch, J. Phys.: Condens.
Matter 10, 2673 (1998)

\bibitem {B181}G. Bergmann, Eur. Phys. J. B 75, 497 (2010)

\bibitem {B177}G. Bergmann, Phys. Rev. B 77, 104401 (2008)

\bibitem {K58}H. R. Krishna-murthy, J. W. Wilkins, and K. G. Wilson, Phys.
Rev. B 21, 1003 (1980)

\bibitem {N14}P. Nozieres, J. Low Temp. Phys. 17, 31 (1974)

\bibitem {A95}P. W. Anderson, Comments in Solid State Phys. 5, 73 (1973)

\bibitem {B182}G. Bergmann, and Y. Tao, Eur. Phys. J. B 73, 95 (2010)
\end{thebibliography}
\end{document}